# Dotted Version Vectors: Logical Clocks for Optimistic Replication

Nuno Preguiça
*CITI/DI*
*FCT, Universidade Nova de Lisboa*
*Monte da Caparica, Portugal*
nmp@di.fct.unl.pt

Carlos Baquero, Paulo Sérgio Almeida,
Victor Fonte, Ricardo Gonçalves
*CCTC/DI*
*Universidade do Minho*
*Braga, Portugal*
{cbm,psa,vff}@di.uminho.pt, rtg@lsd.di.uminho.pt

## Abstract

*In cloud computing environments, a large number of users access data stored in highly available storage systems. To provide good performance to geographically disperse users and allow operation even in the presence of failures or network partitions, these systems often rely on optimistic replication solutions that guarantee only eventual consistency. In this scenario, it is important to be able to accurately and efficiently identify updates executed concurrently. In this paper, first we review, and expose problems with current approaches to causality tracking in optimistic replication: these either lose information about causality or do not scale, as they require replicas to maintain information that grows linearly with the number of clients or updates. Then, we propose a novel solution that fully captures causality while being very concise in that it maintains information that grows linearly only with the number of servers that register updates for a given data element, bounded by the degree of replication.*

## 1. Introduction

The design of Amazon's Dynamo system [1] was an important influence to a new generation of databases, such as Cassandra [2], Riak[1] and Voldemort[2] focusing on partition tolerance, write availability and eventual consistency. The underlying rationale to these systems stems from the observation that when faced with the three conflicting goals of *consistency*, *availability* and *partition-tolerance* only two of those can be achievable in the same system [3], [4]. Facing wide area operation environments where partitions cannot be ruled out, these systems relax consistency requirements to provide high availability.

1. http://www.basho.com/Riak.html
2. http://project-voldemort.com/

The mentioned systems follow a design where the data store is always writable. A consequence is that replicas of the same data item are allowed to diverge, and this divergence should later be repaired. Accurate tracking of concurrent data updates can be achieved by a careful use of well established causality tracking mechanisms [5], [6], [7], [8]. In particular, for data storage systems, version vectors [6] enables the system to compare any pair of replica versions and detect if they are equivalent, concurrent or if one makes the other obsolete. A replica version that is determined to be obsolete can be replaced by a more recent replica version. Merging concurrently modified replicas usually requires *semantic reconciliation*, and this is typically achieved by sending to users (or to higher level application logic) the set of concurrent replica versions, and metadata context, and have them write a new version that supersedes the provided versions.

When accurate causality tracking and handling of concurrent replica versions is considered too complex for a given application domain, systems such as Cassandra resort to physical timestamps derived from node or client clocks, upon which they establish what replica version is considered the most recent. The drawback of this simplification is that it enforces a last writer wins strategy where some concurrent updates are lost. In addition, if the clocks are poorly synchronized some nodes/clients might always lose their competing concurrent updates.

Even in systems where a full-fledged characterization of causality is sought, there are important limitations to either system scalability or to the correctness of the causality tracking in present implementations. In this paper, first we analyze these problems, with a special focus on solutions used in replicated key-value stores designed for cloud computing environments, such as Dynamo, Cassandra and Riak.

We then propose a novel mechanism, *dotted version*

*vectors*, that can provide an accurate and scalable solution to track causality of updates performed by clients. Our approach builds on version vectors; however, unlike previous proposals, it does not require an entry per-client, but only an entry per-server that stores a replica; i.e., according to the degree of replication.

The remainder of this paper is organized as follows. Section 2 presents the system model. Section 3 surveys the current solutions used for causality tracking and discusses their limitations. Section 4 introduces a new kernel of operations on which causality tracking can be based, and on which our proposal is built. Section 5 presents the new mechanism: dotted version vectors. Section 6 discusses related work and Section 7 concludes the paper with some final remarks.

## 2. System model

Storage systems for cloud computing environments can be seen as composed of a set of interconnected server nodes that provide a data read/write service to a much larger set of clients. Without loss of generality, we can consider a standard key-value store interface that exposes two operations: GET(K) and PUT(K,V). (A delete operation can be implemented, for example, by executing a put with a special value.)

A given key is replicated in only a subset of the server nodes, which we call the replica nodes for that key. For our analysis, the approach used to decide which nodes will replicate a given key (e.g., consistent hashing) in not important. Depending on the system, in each replica node, for each key, the system maintains either a single value or multiple concurrent values. We name each of these values, a replica version or simply a version when no confusion may arise.

These systems usually rely on an optimistic replication approach [9], allowing client operations to complete without coordination. In case of concurrent updates to the same key, these systems usually guarantee eventual consistency by either relying on a last writer-wins strategy (e.g. Cassandra [2]) or by maintaining multiple versions for the concurrent updates to the key (e.g. Amazon's Dynamo [1], Depot [10], Riak). In the latter case, conflicts can be solved by issuing a new update that supersedes the concurrent versions, which is usually done by the client (but could also be automatically executed by application code running on a server).

For achieving this execution model, these systems must include some form of causality tracking. In the next section, we analyze the main approaches used currently in these systems, and discuss their properties and limitations.

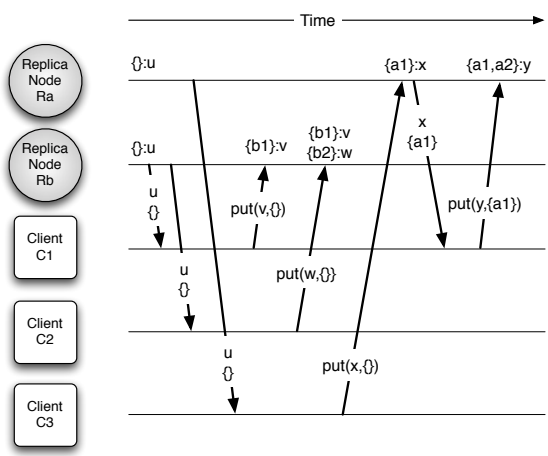

Figure 1. Three clients concurrently modifying the same key on two replica nodes. Causal histories.

An important aspect to consider when reasoning about the scalability of these approaches, is the existence of three different orders of magnitude at play: a small number of replica nodes for each key; a large number of server nodes; a huge number of clients, keys and issued operations. Thus, a scalable solution should avoid mechanisms that are linear with the highest magnitude and, if possible, even try to match the lowest scale.

## 3. Common approaches to Causality tracking

When considering the system composed by the clients and the storage system, a large number of causal relations are established as the clients issue operations to the servers. Different key-value storage systems trace different sub-sets of these relations.

One simple way to formally illustrate this is to use *causal histories* [7]. Causal histories are simply described by sets of unique update event identifiers. These unique update identifiers can be generated by a unique node identifier and a monotonic integer counter. (We will use replica-based identifiers but client identifiers could be used as well. The crucial point is that identifiers have to be globally unique.) The partial order of causality can be precisely tracked by comparing these sets by set inclusion. Two histories are concurrent if neither include the other: $A \parallel B$ iff $A \nsubseteq B$ and $B \nsubseteq A$.

Consider a simple example, illustrated in Figure 1: Clients $C_1, C_2, C_3$ read the same state from synchronized replica nodes and do independent updates. In this simplified description we omit the keys, implicitly assuming they are the same, and only show the causal information that is committed to each replica node and respective versions, in the same order.

When client $C_1$ does its first PUT, replica node $R_b$ will record the version associated with the causal history $\{b_1\}$ that includes the update identifier, $b_1$, and the history previously observed by $C_1$, $\{\}$ in this case. When client $C_2$ does its PUT, the causal history associated with the new value will be $\{b_2\}$, which does not include $b_1$ because $C_2$ has not observed this version. Thus, $R_b$ ends up with two concurrent versions, as stated by the causal histories. The second PUT from client $C_1$, handled by $R_a$, supplies a new version $y$ together with its knowledge of causal history $\{a_1\}$, obtained from its last GET. Replica $R_a$ records this update and adds $a_2$ to its corresponding causal history. Since $\{a_1\} \subset \{a_1, a_2\}$ the version $y$ will syntactically dominate $x$ and replace it in the committed state in $R_a$.

At the end of the run we have a value $y$ in $R_a$ than can be detected, by the causal histories, to be concurrent with the two concurrent values, $v$ and $w$, stored on replica node $R_b$.

This very simple model assumes that the client maintains no state other than the context of the last GET when executing a PUT operation. This may lead to unexpected results for a client issuing a sequence of operations. For example, a client, after observing some given version of the data, may later observe an older version. To address this problem, the client could maintain the causal history of observed data, which would contain all the update identifiers observed. This could be easily computed by the union of the causal histories returned in GET operations and the identifiers of executed PUT operations (which could be returned to the client when acknowledging the PUT). This causal history could be passed as an argument in GET operations, restricting the servers that could process the operation to the ones that contain the given causal history (e.g., Bayou provides these session guarantees, using a solution based on version vectors [11]).

Although conceptually simple, causal histories are not adequate for use in practical systems, since they scale linearly with the number of updates. Next, we survey the mechanisms used in actual systems.

### 3.1. Causally compliant total order

One simple approach is to establish a total order among updates that is compliant with causal dependencies, and use this order to enforce a *last writer wins* policy. The simplest total order is obtained assuming that client clocks are well synchronized and applying real time clock order (simultaneous events are usually further ordered over process ids). In this approach, replica nodes never store multiple versions and writes do not need to provide a get context.

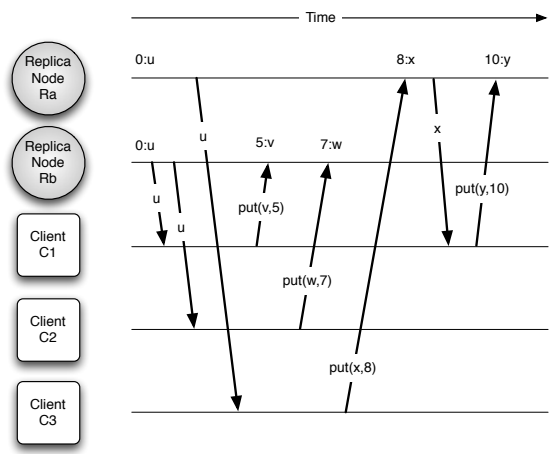

Figure 2. Three clients concurrently modifying the same key on two replica nodes. Perfectly synchronized real time clocks.

Figure 2 depicts the same run used to illustrate the use of causal histories, but now using perfectly synchronized client clocks. One can observe that concurrent events are ordered by the clocks and that the total order is compliant with the causal order: If two values would have causal histories $c$ and $c'$ such that $c \subset c'$ then the real time clocks $t$ and $t'$ are such that $t < t'$. This can be verified, observing values $x$ and $y$ in the run.

The problem is that although causally we have a partial order with $\{a_1, a_2\} \parallel \{b_1\} \parallel \{b_2\}$, this approach ends up ordering all updates. The total order established is compliant with causality, but will order actions that are in fact concurrent.

The approach based on client real time clocks is used in Cassandra v0.6.x (and v0.7.0 betas) [2], and referred in Dynamo as an alternative to version vectors for some application settings.

An important drawback with real time is that if client clocks go out of sync the total order might no longer be compliant with causality. It is easy to see that a client with systematically delayed clock values will never see its updates committed and, conversely, that if a clock is always advanced its client updates will always win over concurrent ones.

An alternative approach that avoids real time clock synchronization and the potential anomalies when it fails, would be to use Lamport clocks [5], establishing a total order among updates that is compliant with causal dependencies. This can be easily achieved by assigning a pair (CLOCK,REPLICA), where CLOCK is a Lamport clock and REPLICA, a unique site identifier that can be either the identifier of the client or the replica node that received the update. As usual, for tracing causality, the local clock used to tag new updates must be updated when the client gets a newer version of the data. A total order is established on these pairs, as usual, with $(c_a, r_a) < (c_b, r_b)$ iff $c_a < c_b \vee (c_a = c_b \wedge r_a < r_b)$. Again, this total order would not represent concurrent events.

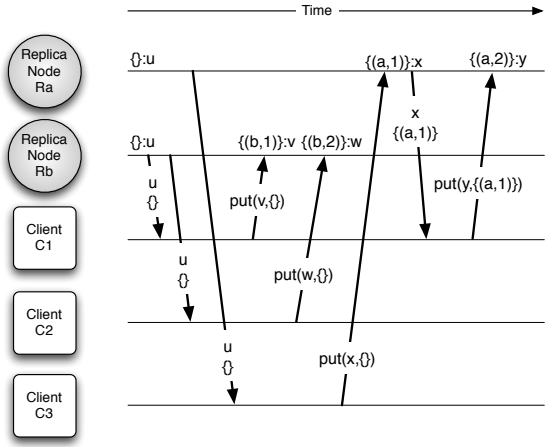

Figure 3. Three clients concurrently modifying the same key on two replica nodes. Per-server entries.

### 3.2. Version vectors with per-server entry

A second approach is to track causality by using *version vectors* [6] with an entry per server replica node. In this case, each server maintains a version vector where each entry summarizes the sequence of updates it reflects. For example, a causal history of sequential replica events $\{a_1, a_2, b_1, b_2, c_1\}$ is summarized as $\{(a,2), (b,2), (c,1)\}$. In traditional version vectors, for a fixed and ordered set of nodes, this can be further summarized as $[2, 2, 1]$, but this notation is not adequate for dynamic systems where the number of nodes can vary over time.

When the client executes a GET operation, it receives the version vector summarizing the causal history of events reflected in the version(s) received. Later, when the client executes a PUT, it sends the context on which the update is executed, i.e., the version vector previously received. The replica node increments its local counter to reflect the new update, and stores it in the entry of the received vector corresponding to its own identifier. It then checks if this new vector causally dominates any version currently stored, and discards any version made obsolete.

In this case, it is possible to track the causality among updates that were received in different servers. Figure 3 depicts the same example run but now (as opposed to Figure 2) updates $y$ and $w$ are correctly detected to be concurrent, since $\{(a,2)\} \parallel \{(b,2)\}$. If a client GET collects these versions from the two replica nodes, this concurrency will be exposed and the client, receiving two versions, can submit back a version that dominates both updates.

However, this approach cannot track causality among updates submitted to the same server. In the example, when the update $w$ from the client $C_2$ is submitted to replica $Rb$, it will get registered with the version vector $\{(b,2)\}$ and appear to dominate the previous committed value $v$ with vector $\{(b,1)\}$. By comparing the version vectors of both updates, they will not be considered concurrent. This can be surprising considering the fact that if the second client had submitted the update to a different server it would be considered concurrent.

In practice a last writer wins policy was enforced with respect to concurrent updates handled in the same replica node, and, in this case, one concurrent update was lost. This linearization of concurrent updates, due to the use of less version vector entries than sources of concurrent activity, is formalized in *plausible clocks* [12]. The Dynamo system uses one entry per replica node and thus falls into this category.

The reason for the concurrent updates of the two clients submitted to the same server not being considered concurrent is consequence of the fact that the version vector associated with the second update does not correctly summarizes its causal history. In fact, the vector $\{(b,2)\}$ summarizes updates $\{b_1, b_2\}$, which includes the update $v$ of the first client $C_1$. One can argue that the replica node $Rb$ could instead verify that the new update is concurrent with its current version by checking that the version vector included in the operation does not dominate the version vector of the current version. In this case, the replica node could reject the update, implementing a conditional write semantics. This approach is used, e.g., in Coda [13] and in the CVS version control system (although not necessarily relying on version vectors). However this goes against the usual policy of write availability [1]. The other possibility would be to register the conflict and maintain both data versions. In this case, the problem is that there is no version vector the replica node $Rb$ could generate that traces the dependency with the other version, as $\{(b,2)\}$ would be interpreted as overwriting $\{(b,1)\}$.

### 3.3. Version vectors with per-client entry

We have seen that version vectors with one entry per replica node are not enough to track causality among concurrent clients[3]. One natural approach is to track causality by using version vectors with one entry per client (if servers can also update the data, with server side scripts, an entry for each server should also be included in the version vector). Now the number of entries matches the number of concurrency sources and one no longer faces a plausible clocks setting.

As in the previous approach, updates are associated with a version vector. When a client executes a GET operation, it will receive the version vector associated with the version(s) that it reads. Later, when a client submits a new update, using PUT, the replica node will receive this vector and the client identity.

With per-client entries, the correct way to obtain the integer value used to register the update would be for each client to maintain a counter, increment it and provide it in each PUT operation, together with the context previously received. A version vector for the new version can be obtained from the context version vector by replacing the entry of the client by the given value. If we want to support a model with stateless clients, which only provide the context received by a GET and their unique identifier, we can do so if we have a *read your writes* semantics [11] (obtained, e.g., through read and write quorums), so that the most recent update by a given client is present in the context.

Otherwise, the server can, at most, try to infer the moat recent update by that client, by using the maximum of the respective entry in the received context and all vectors at the server for that key. As a more recent update by that client can be stored in other server, this can lead to lost updates.

In Figure 4, in the usual run, we illustrate this problem. Client $C_1$ when writing $v$ in node $Rb$, has its updated registered as $(C_1, 1)$. Its later updates will get distinct, and monotonically increasing values as long as the client reads its last written version. However in this run, the client will issue a later update in replica $Ra$ and this update will again be registered with $(C_1, 1)$. The consequence is that now value $v$ seems to be dominated by version $y$, since $\{(C_1, 1)\} < \{(C_1, 1), (C_3, 1)\}$.

---

3. An interesting discussion on this issue can be seen in http://blog.basho.com/2010/04/05/why-vector-clocks-are-hard/

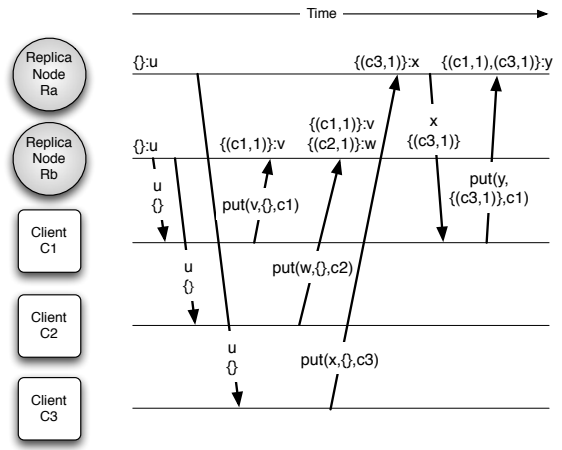

Figure 4. Three clients concurrently modifying the same key on two replica nodes. Per-client entries.

Although, when used correctly, this approach can fully trace the causality among concurrent updates submitted by different clients, it has the obvious drawback of requiring one entry per client, which makes the size of the vectors now linear with the number of clients that perform PUT operations.

Many web and cloud computing systems are based on a three-tier architecture, where the clients of the storage system are application servers (or other middleware components). Although this seems to alleviate the scalability problem, as the number of these components is small when compared to the number of end-clients, this is not the case. In fact, as these application servers (or middleware components) run concurrent threads of activity, it is necessary to have an entry for each thread of activity or concurrent updates in different threads would not be considered concurrent (as in the per-server entry approach). As threads of activity are very dynamic, this also poses the problem of how to keep track of their identifiers.

## 4. A Kernel for Eventual Consistency

We have seen that, as soon as clients can perform concurrent updates managed by a single replica node, several concurrent versions may result, that have to be kept in that node. These version sets are returned by a get operation, and their clocks are supplied as the context in a put operation.

In this section we argue that the mechanics of a distributed key-value store, in terms of causality tracking, should be based on two core functions on the sets of logical clocks of replicas.

- sync$(S_1, S_2)$: takes two clock sets and returns a clock set. It returns a set of concurrent clocks, each belonging to one of the sets, and that together cover both sets while discarding obsolete knowledge;
- update$(S, S_r, r)$: takes a clock set ($S$, the context supplied by the client), the set of clocks in the replica node $S_r$, and the replica node id $r$, and returns a clock. This clock should dominate all clocks in $S$ and not be dominated by any join of clocks in the system.

More formally, in a given system containing the replica nodes $R$, each with a version set $S_i$, with $i \in R$, these operations should be defined in a way as to respect the following conditions. (Where we use $\sqcup$ for the join operation on clocks, assuming as usual that the partial order on clocks is a join semilattice.)

- If $S = \text{sync}(S_1, S_2)$, then:
  1) $\forall x \in S.\, x \in S_1 \cup S_2$,
  2) $\forall x, y \in S.\, x \not\leq y$,
  3) $\forall x \in S_1 \cup S_2.\, \exists y \in S.\, x \leq y$.
- If $u = \text{update}(S, S_r, r)$, then:
  1) $\forall x \in S.\, x \leq u$,
  2) $\forall x \in \bigcup_{i \in R} S_i.\, (x \leq u \Rightarrow x \leq \bigsqcup S)$,
  3) $u \not\leq \bigsqcup \bigcup_{i \in R} S_i$.

The function sync produces a set of concurrent clocks that describe the collective causal past in the parameters. It simply returns elements from the sets in the parameters, and it can have a general implementation, defined only in terms of the partial order on clocks, regardless of their actual representation:

$$\text{sync}(S_1, S_2) = \{x \in S_1 \mid \nexists y \in S_2.\, x < y\} \cup \{x \in S_2 \mid \nexists y \in S_1.\, x < y\}$$

The update operation can be more of a challenge because its constraints involve a global condition on the system, but it must be implemented without global knowledge. This is specially the case in dynamic systems, as described in [14], but here we have the challenge of how to avoid the use of client identifiers in clocks. Also, update needs to generate values that are not in the sets of clocks passed as parameter, and that will depend on the concrete representation of clocks.

Before developing some implementation, it is useful to characterize clock mechanisms using causal histories as a reference, as in Section 3.

In term of causal histories, the update function is:

$$update(S, S_r, r) = \bigcup_{X \in S} X \cup \{e\} \qquad \text{with } e \notin \bigcup_{i \in R} \bigcup S_i$$

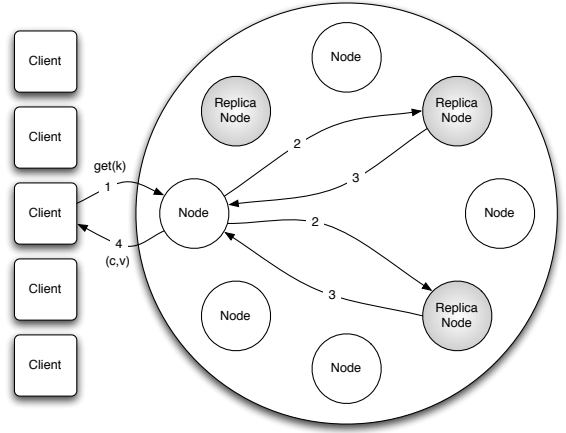

Figure 5. A get operation.

which resorts to an oracle with global knowledge to obtain a globally unique event identifier. (As shown in Figure 1, these event names can be obtained with unique replica node identifiers and monotonic counters.)

### 4.1. Using the kernel operations

A key-value store can now implement the operations it intends to make available to clients by using the kernel operations sync and update.

**Operation** get$(k)$. When a client asks some proxy node $P$ to perform a get of some key $k$ (step 1 in Figure 5):
- $P$ computes the set of replica nodes $R$ for $k$;
- $P$ ask to a subset of nodes in $R$ for the value for that key. Depending on the expected semantics, this subset may contain, for example, a single node or a quorum of nodes (step 2);
- $P$ waits for the replies (step 3);
- $P$ performs a reduce of the replies using the sync operation, and replies to the client (step 4).

**Operation** put$(k, v, S)$. When a client asks some proxy node $P$ to perform a put for some key (step 1 in Figure 6):
- $P$ computes the set of replica nodes $R$ for $k$;
- if $P$ is a replica node for $k$, then $P$ will coordinate the request; otherwise $P$ will forward the request to some replica node for $k$, that will act as coordinator (step 2);
- the coordinator $C$ performs an update operation, resulting in a clock value $u = \text{update}(S, S_C, C)$;

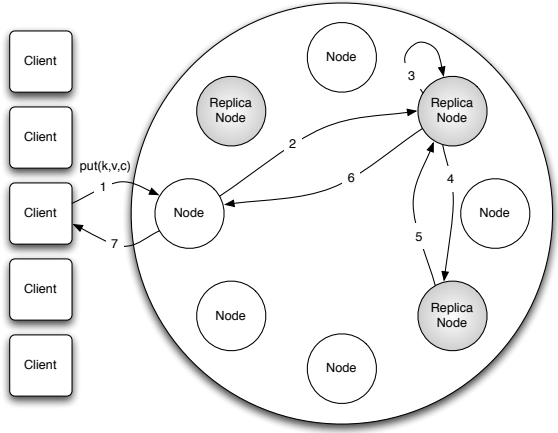

Figure 6. A put operation.

- $C$ performs a sync between $u$ and the local set of instances, and stores the result of the sync $S'_C = \text{sync}(S_C, \{u\})$ (step 3);
- $C$ sends $S'_C$ to a subset of other nodes in $R$. Depending on the expected semantics, this subset may, for example, be empty or contain a quorum of nodes (step 4);
- each of those nodes performs a sync between $S'_C$ and the local set of instances, stores the result of the sync $S'_i = \text{sync}(S_i, S'_C)$, and acknowledges to $C$;
- $C$ waits for the replies (if the subset is not empty) (step 5);
- $C$ acknowledges to the proxy $P$ (step 6), which in turn acknowledges to the client (or $C$ acknowledges directly if that is possible) (step 7).

**Anti-entropy.** In addition to what is done in the above operations, nodes can at any moment decide to engage in anti-entropy. A replica node can send its state, including the clock set and version to other replica node. The receiving node performs a sync with the local entry for that key, and stores the result locally.

## 5. Dotted Version Vectors

We now present a concise and accurate representation for the clocks to be used as a substitute for the classic version vectors in key-value stores. The mechanism allows a lossless representation of causality (contrary to, e.g., Plausible Clocks) while only using server-based ids, and only a component per replica node, thus avoiding the space consumption explosion that occurs in id-per-client approaches.

While a version vector compresses causal histories by representing, for each component, all events in a range up to a given sequence number, we will be able to represent also individual events that fall outside such ranges.

As an example, a version vector $\{(a, 2), (b, 1), (c, 3)\}$ represents the causal history:

$$\{a_1, a_2, b_1, c_1, c_2, c_3\}.$$

We will be able to represent a causal history like:

$$\{a_1, a_2, b_1, c_1, c_2, c_3, c_7\},$$

where event $c_7$ falls outside the range from 1 to 3.

Dotted version vectors are able to represent, for any given component, both a range, and a range plus and individual event (a "dot"). We will see that a range plus a single event (as opposed to arbitrary sets) is enough for the scenario at hand.

### 5.1. Definition

A dotted version vector is a logical clock which consists of a mapping from identifiers to either integers or pairs of integers $(m, n)$. For notational convenience we will use instead a triple $(id, m, n)$ for such elements of the mapping. The events represented by a clock can be characterized by a semantic function from clocks (or sets of clocks) to causal histories:

$$\begin{aligned}
\mathcal{C}[\![(r, m)]\!] &= \{r_i \mid 1 \leq i \leq m\}, \\
\mathcal{C}[\![(r, m, n)]\!] &= \{r_i \mid 1 \leq i \leq m\} \cup \{r_n\}, \\
\mathcal{C}[\![X]\!] &= \bigcup_{x \in X} \mathcal{C}[\![x]\!].
\end{aligned}$$

In a component $(r, m, n)$ we will always have $n > m$.

With this definition, the causal history:

$$\{a_1, a_2, b_1, c_1, c_2, c_3, c_7\},$$

that cannot be represented in a version vector, will be represented by the dotted version vector $\{(a, 2), (b, 1), (c, 3, 7)\}$.

### 5.2. Partial order

The order on clocks can be defined, as usual, in terms of inclusion of causal histories; i.e.:

$$X \leq Y \iff \mathcal{C}[\![X]\!] \subseteq \mathcal{C}[\![Y]\!]$$

This can be computed by the function on mappings:

$$X \leq Y \iff \forall x \in X.\, \exists y \in Y.\, x \leq y,$$

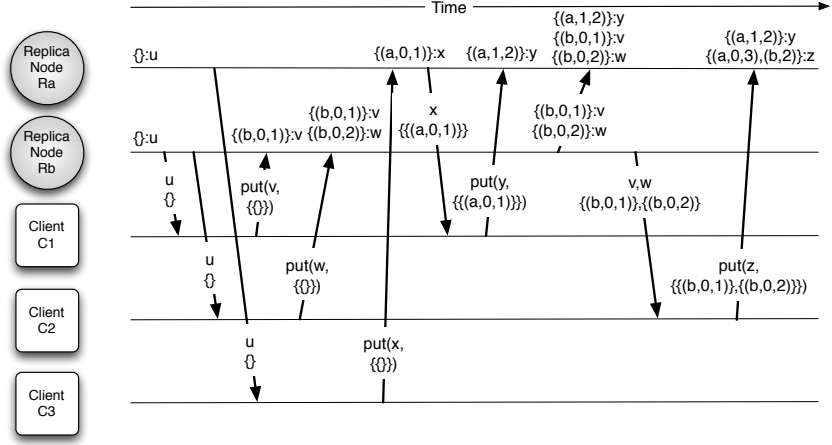

Figure 7. Three clients concurrently modifying the same key on two replica nodes. Dotted version vectors.

where the order on individual components of the mapping is defined by the clauses:

$$\begin{array}{rcll}
(r,m) & \leq & (r,m') & \text{if } m \leq m', \\
(r,m) & \leq & (r,m',n') & \text{if } m \leq m' \vee m = m'+1 = n' \\
(r,m,n) & \leq & (r,m') & \text{if } n \leq m', \\
(r,m,n) & \leq & (r,m',n') & \text{if } n \leq m' \vee (m \leq m' \wedge n = n'), \\
x & \not\leq & y & \text{otherwise.}
\end{array}$$

This order allows concurrent clocks even using only a component from a single replica node. As an example:

$$\{(r,4)\} \parallel \{(r,3,5)\},$$

as they represent the causal histories:

$$\{r_1, r_2, r_3, r_4\} \parallel \{r_1, r_2, r_3, r_5\},$$

This situation will arise when $\{(r,4)\}$ is stored in a replica node and a client, which in the past has read some value and got the context $\{(r,3)\}$, now performs a put using this context. This situation is very common but cannot be handled with current mechanisms using server-based identifiers.

In Figure 7, we present our usual run using dotted version vectors. It can be seen that causality is accurately tracked, even tough per-server identifiers are used. We also extend the run so that replica node $R_b$ decides to do some anti-entropy and sends state to node $R_a$ that syncs its. Then, client $C_2$ does an interaction (with no affinity) where it reads from $R_b$ and does an update $z$ to $R_a$. We can see that, as expected, $z$ will subsume both $v$ and $w$, and is registered as concurrent to $y$.

**5.3. Update function**

An update registered on a replica node $r$ containing the set of versions $S_r$, can have a reference definition, in terms of causal histories using replica node ids plus sequence numbers to distinguish events, as:

$$\text{update}(S, S_r, r) = \bigcup S \cup \{r_{n+1}\} \quad \text{with}$$
$$n = \max(\{0\} \cup \{x \mid r_x \in \bigcup S_r\}).$$

To define the update function, we make use of some auxiliary functions. The ids function gives the set of identifiers in a clock or set of clocks:

$$\begin{array}{rcl}
\text{ids}((r,\_)) & = & r, \\
\text{ids}((r,\_,\_)) & = & r, \\
\text{ids}(X) & = & \{\text{ids}(x) \mid x \in X\}.
\end{array}$$

The $\lceil \_ \rceil_\_$ function takes a clock or set of clocks and a replica node identifier and returns the maximum integer contained in the mapping from that identifier:

$$\begin{array}{rcl}
\lceil C \rceil_r & = & \max(\{0\} \cup \{m \mid (r,m) \in C \vee (r,\_,m) \in C\}), \\
\lceil S \rceil_r & = & \max(\{0\} \cup \{\lceil C \rceil_r \mid C \in S\}).
\end{array}$$

The update function can now be defined:

$$\begin{array}{rcl}
\text{update}(S, S_r, r) & = & \{(i, \lceil S \rceil_i) \mid i \in \text{ids}(S) \wedge i \neq r\} \cup \\
& & \{(r, \lceil S \rceil_r, \lceil S_r \rceil_r + 1)\}.
\end{array}$$

It can be seen by this definition that (given that sync does not generate new values) all clocks have exactly one component which is a triple; all the others are the same as in classic version vectors. This means that dotted version vectors can also be thought of as a standard version vector augmented by a pair identifier-counter to describe the single dot needed.

In the example of Figure 7, each put operation generates a new clock for the new version. The first PUT from client $C_1$ generate the clock $(b, 0, 1)$, as no version exists previously in replica node $Rb$. The same for the first PUT from client $C_3$ on replica node $Ra$, which generates the clock $(a, 0, 1)$. A more interesting case is the first PUT from client $C_2$ on replica node $Rb$. In this case, as there is a version in replica node $Rb$ with a clock that is not dominated by the context of the PUT, $\{\}$, the clock generated is $(b, 0, 2)$, encoding only the event $b_2$ of the causal history.

The second PUT from client $C_1$ exemplifies the situation where a client overwrites the version it has previously read. In this case, the generated clock is $(a, 1, 2)$, as the read context dominates (is equal in this case) to the clock of the version in the replica node.

The most complex example arises in the second PUT from client $C_1$. This example exemplifies the situation where a client receives two concurrent versions and creates a new version that superseeds the previous concurrent updates. In this case, the context of the PUT is $\{\{(b, 0, 1)\}, \{\{(b, 0, 2)\}\}\}$, and the clock generate in replica node $Ra$ is $\{(a, 0, 3), (b, 2)\}$. The component $(b, 2)$ encodes the events $b_1, b_2$ of the causal history, which were represented in the context of the PUT. The component $(a, 0, 3)$ registers the new update event $a_3$ associated with this PUT operation.

### 5.4. Correctness

The operations on a key-value store invoked by clients (get and put) resort to the kernel operations sync and update. These operate on (and return) sets of clocks. Single clocks are not a first class entity that can be operated upon by clients. A client may perform a get, which may return a set of concurrent replicas and the opaque context for the corresponding set of clocks. The client may use the context on a subsequent put operation, but cannot operate upon individual clocks from that context.

The reason that makes it possible to have an accurate representation of causality using dotted clocks, is that all sets of clocks that are kept at replica nodes or returned to clients have the invariant that, for each node identifiers present in (some element of) the set, all sequence numbers from 1 up to some given value will be present in the union of the corresponding causal histories.

More formally, we define the predicate over clock sets:

$$\text{downset}(S) \iff \forall i \in \text{ids}(S). \forall\, 1 \leq n \leq \lceil S \rceil_i. i_n \in \mathcal{C}[\![S]\!],$$

which is true for sets of clocks for which the union of the corresponding causal histories are downward closed sets under the order over events $r_i \leq s_j \iff r = s \wedge i \leq j$. In other words, the predicate is true if, for each node $r$, the set contains all events generated by $r$ up to some given point in time.

We now show that, in a given system containing replica nodes $R$, each $r \in R$ with a replica set $S_r$, the following invariant holds:

$$\forall r \in R.\ \text{downset}(S_r).$$

It is easy to see that if both $X$ and $Y$ are downsets, then $Z = \text{sync}(X, Y)$ will also be a downset. This means that if we assume that the invariant holds, then the values returned to clients are downsets. It also means that, if $S'_C$ computed by the coordinator in a put operation is a downset, then the values stored in other replica nodes after receiving a store request from the coordinator will also be downsets.

Therefore, to prove that the invariant holds, the only interesting case is what happens in the coordinator in a put operation. The new version set to be stored locally will the result of an update followed by a sync: $u = \text{update}(S, S_C, C)$ and $S'_C = \text{sync}(S_C, \{u\})$. Assuming that $S$ and $S_C$ are downsets, $S'_C$ will also be a downset because:

- $u$ will be present in the result $S'_C$;
- although $\{u\}$ itself may not be a downset, for any identifier $i$ other than $C$, the computed mapping $(i, \lceil S \rceil_i)$ represents a contiguous range of events starting from 1 for identifier $i$ in the corresponding causal history of $S$; the sync between $\{u\}$ and $S_C$ will therefore be a downset in what concerns these identifiers;
- for identifier $C$, as some clock $v$ in $S_C$ will not be present in $S'_C$ only if $v < u$, as $S_C$ represents all events from $C$ up to $\lceil S_C \rceil_C$, and $u$ contains only one more event with number $\lceil S_C \rceil_C + 1$, then $S'_C$ represents a contiguous range starting from 1 for id $C$.

To summarize: even though $\{u\}$ is not necessarily a downset, syncing it with the clock set in the coordinator will result in a downset, as only a successor event is added and no "holes" are created.

It remains to point out that, as the clock set $S$ sent as context from the client is a downset, the clock $u$ computed in the update can represent the appropriate causal history (the union of the causal histories corresponding to clocks in $S$ plus a new event) accurately, with no loss of information.

## 6. Related Work

Version tracking solutions as used in cloud storage systems are rooted on Lamport' seminal work on the definition and role of causality in distributed systems [5]. This work was the foundation for subsequent advances in causality's basic mechanisms and theory, including the introduction of version vectors [6] for tracking causality among replicas in a distributed storage system and vector clocks [15], [16] for tracking causality of events in a distributed systems.

Most of this initial work dealt with a fixed, mostly small, number of participants. Later, several systems introduced mechanisms for the dynamic creation and retirement of vector entries to be used when a server enters and leaves the system. While some techniques required the communication with several other servers [17], others required communication with a single server [18]. Interval Tree Clocks [14] are able to track causality in a dynamic, descentralized scenario where entites can be autonomously created and retired. Other systems, such as Dynamo [1], use unsafe techniques to remove entries that are expected not to be necessary based on time.

Even with these mechanisms, tracking causality through version vectors or vector clocks requires a space linear with the number of entities in the system, posing scalability problems for system with a large number of elements [19]. This problem is experienced in practice, for example, in cloud computing storage systems, as discussed in Section 3.

Besides the safe techniques previously mentioned to remove entries that are no longer needed, several other directions have been tackled to address this problem.

The Roam system [20] runs a consensus protocol to decrease, in all servers, the value of all entries of the version vector by a constant value. The system only keeps the entries that are larger than zero. The *dependency sequences* [21] mechanism assumes a scenario where dynamic, weakly-connected sets of entities (mobile hosts) communicate through designated proxy entities chosen from a stable, well-connected (mobile service stations). The mechanism maintains information about the causal predecessors of each event. It needs to take periodic global snapshots to prune discardable causality-tracking metadata.

In Depot [10], the version vector associated with each update only includes the entries that have changed since the previous update in the same node. However, each node still needs to maintain version vectors that include entries for all clients and servers. In a similar scenario, the same approach could be used as a complement to our solution.

Other storage systems explore the fact that they manage a large number of objects to maintain less information for each object. In Microsoft's WinFS [22], a base version vector for all objects is maintained for the file system, and each object maintains only the difference for the base in a concise version vector. In Cimbiosys [23], the authors suggest the use of the same technique in a peer-to-peer system. These systems, as they maintains only one entry per server, cannot generate two concurrent version vectors for tagging concurrent updates submitted to the same server from different clients, as discussed in Section 3.

Another direction is to use unsafe space-folding approaches that can reduce the storage and communication overhead at the expense of less accuracy of the causality relation captured by these mechanisms. Although devised as an alternative not to version vectors but to vector clocks, *plausible clocks* [12] propose techniques for condensing event counting from multiple replicas over the same vector entry. The resulting order does not contradict the causal precedence relation but because counters are effectively shared between processes, some concurrent events will be perceived as causally related. In fact, the previously mentioned Lamport clocks [5], are a notable example of plausible clocks.

Another approach trading off less accuracy of causality-tracking for better scalability is the *hash history* mechanism [24]. It provides a directed graph not of update operations, but of version hashes over the state of each replica. Although independent of the number of replicas in the system, the storage overhead grows linearly with the number of updates. In order to minimize this problem, it truncates the histories, pruning the oldest hashes based on loosely synchronized clocks. Use of hashes, however, can only guarantee statistical correctness, and pruning may cause incorrect perception of concurrency.

## 7. Conclusion

In this paper we have introduced *dotted version vectors*, a novel solution for tracking causal dependencies among update events. The base idea of our solution is to efficiently encode the causal history of an update performed by some client, as a version vector that encodes the causal history of the previously observed state and a single event identifier assigned in the server that receives the update.

*Dotted version vectors* allow to accurately track causality among updates executed by multiple clients using information that is only linear with the number of servers that register these updates, bounded by the degree of replication. When compared with previously proposed safe solutions that require information linear with the number of clients, our solution is much more efficient, as the number of clients tends to be several orders of magnitude larger than the number of servers that register updates for a given data element.

As an evidence that dotted version vectors can be easily used in current cloud storage systems, we have modified the Riak key-value store to integrate our solution. Besides the definition of dotted version vectors, approximately 100 lines of code had to be modified.

In the future, we intend to study how the underlying idea of dotted version vectors can be applied to other mechanisms to track causality, such as extending Interval Tree Clocks [14] in order to better handle membership changes in the set of replica nodes.

## 8. Acknowledgment